\documentclass[12pt]{article}
\usepackage[utf8]{inputenc}
\usepackage{hyperref}
\usepackage{amsmath}
\usepackage{graphicx}
\usepackage{enumitem}
\usepackage{algorithm}
\usepackage{algpseudocode}
\usepackage{listings}
\usepackage{xcolor}
\usepackage{float}
\usepackage{subcaption}

\lstdefinelanguage{ccsjson}{
  basicstyle=\ttfamily\footnotesize,
  keywordstyle=\color{black},
  stringstyle=\color{black},
  commentstyle=\color{gray},
  showstringspaces=false,
  breaklines=true,
  frame=single,
  columns=fullflexible
}

\usepackage{float}  % Add this in your preamble

\usepackage{listings}
\usepackage{xcolor}

\title{\vspace{-2cm}AI Agents Need Memory Control Over More Context\vspace{1cm}}

\author{
    Fouad Bousetouane\\[1em]
    \text{The University of Chicago, USA}\\
    {\small \href{mailto:bousetouane@uchicago.edu}{\texttt{bousetouane@uchicago.edu}}}
}

\date{}

\begin{document}

\maketitle

\begin{abstract}
AI agents are increasingly used in long, multi-turn workflows in both research and enterprise settings. As interactions grow, agent behavior often degrades due to loss of constraint focus, error accumulation, and memory-induced drift. This problem is especially visible in real-world deployments where context evolves, distractions are introduced, and decisions must remain consistent over time.

A common practice is to equip agents with persistent memory through transcript replay or retrieval-based mechanisms. While convenient, these approaches introduce unbounded context growth and are vulnerable to noisy recall and memory poisoning, leading to unstable behavior and increased drift.

In this work, we introduce the \emph{Agent Cognitive Compressor (ACC)}, a bio-inspired memory controller that replaces transcript replay with a bounded internal state updated online at each turn. ACC separates artifact recall from state commitment, enabling stable conditioning while preventing unverified content from becoming persistent memory.

We evaluate ACC using an agent-judge-driven live evaluation framework that measures both task outcomes and memory-driven anomalies across extended interactions. Across scenarios spanning IT operations, cybersecurity response, and healthcare workflows, ACC consistently maintains bounded memory and exhibits more stable multi-turn behavior, with significantly lower hallucination and drift than transcript replay and retrieval-based agents. These results show that cognitive compression provides a practical and effective foundation for reliable memory control in long-horizon AI agents.
\end{abstract}

\newpage
\tableofcontents
\newpage

\section{Introduction}
\label{sec:intro}

Agentic AI is moving beyond conversational assistance toward operational decision support. In enterprise settings, AI agents are increasingly expected to execute multi-step workflows, coordinate external tools, and sustain context across extended multi-turn interactions in domains such as IT operations, cybersecurity response, healthcare operations, and e-commerce. These settings differ from short-form question answering because success depends on continuity under changing requirements, reliable preservation of constraints, and consistent tracking of entities and intermediate decisions. A widely adopted execution pattern is to interleave reasoning and acting during runtime \cite{yao2022react}.

Despite rapid progress in agent architectures, memory handling remains a central barrier to reliability in multi-turn workflows. The dominant implementation pattern continues to rely on transcript replay, where prior interactions are appended to the prompt. As interactions extend, this strategy inflates cost and latency, reduces selectivity by forcing attention over increasingly mixed history, and makes early mistakes harder to recover from. In practice, this presents as drift from established constraints, inconsistent adherence to required formats, and the carryover of unsupported assumptions.

Retrieval-based memory partially addresses prompt growth by storing prior interactions in an external store and retrieving top-ranked artifacts per turn \cite{lewis2020rag,karpukhin2020dpr}. Memory orchestration systems extend this idea by managing what stays in the prompt versus what is stored externally \cite{packer2023memgpt}. Reflection-based approaches store distilled lessons or self-critique to influence later turns \cite{shinn2023reflexion}. These methods are useful, but they leave an important gap for operational settings. Retrieved or summarized text is not equivalent to a controlled internal state. Retrieval optimizes for relevance signals such as semantic similarity, not for preserving the currently active constraints, and selection errors can compound across turns. What agents often need to preserve is a small set of invariants such as goals, policy constraints, service-level constraints, entity identifiers, and confirmed decisions, while continuously integrating new evidence and ignoring distractions.

We frame this as a memory control problem. Sustained multi-turn performance requires an explicit internal state that is updated online at each turn, rather than an uncontrolled accumulation of text. Biological cognition provides a useful analogy. Human working memory is capacity-limited and relies on compression, selection, and executive control \cite{baddeley1974working,cowan2001magical}. Rather than replaying experience verbatim, humans maintain compact task representations that separate recent episodic updates from stable semantic structure \cite{tulving1972episodicsemantic}. This motivates an engineering principle for agentic systems: carry forward a bounded working state that preserves task-critical invariants while filtering noise.

Based on this principle, we introduce the \emph{Agent Cognitive Compressor (ACC)}, a cognitively inspired memory controller that replaces transcript replay with a bounded Compressed Cognitive State (CCS). At each turn, ACC updates CCS using the current interaction, the previous CCS, and a small set of retrieved artifacts. The design explicitly separates artifact recall from state commitment. Retrieval proposes candidate information, while compression commits only what is required for control. CCS then conditions downstream reasoning and action, enabling stable behavior across extended multi-turn interactions with bounded memory growth.

We evaluate the ACC agent against two alternative agents, a transcript replay baseline and a retrieval-based agent, on multi-turn scenarios designed to reflect operational conditions, including evolving constraints, injected distractions, and entity-dependent decisions. Our evaluation uses an agent-judge-driven live framework that issues the same query to all three agents at each turn and produces turn-level scores for task outcome metrics, with bias controls such as blinding and randomized presentation order \cite{zheng2023mtbench,dubois2024length}. To quantify memory-driven anomalies, we use a memory-aware judge that audits hallucination and drift against a judge-maintained canonical state, and we report direct measurements of memory footprint across turns. Results show that the ACC agent improves multi-turn consistency and constraint retention while substantially reducing persistent context size. The remainder of the paper is organized as follows. Section~\ref{sec:background} reviews background and related work on agent architectures and memory. Section~\ref{sec:acc} introduces the Agent Cognitive Compressor (ACC) and the Compressed Cognitive State (CCS), and Section~\ref{subsec:acc_in_architectures} shows how ACC integrates into common agent architecture patterns. Section~\ref{sec:evaluation} presents the agent-judge-driven multi-turn evaluation framework and reports the empirical results. Section~\ref{sec:conclusion} concludes with implications and directions for future work.

\section{Background and Related Work}
\label{sec:background}

Modern AI agents extend large language models with an orchestration layer that manages control flow, tool invocation, and state across multi-step execution. This section reviews common agent patterns and the memory mechanisms used in practice, then motivates cognitive compression as a principle for stabilizing multi-turn agent behavior.

\subsection{AI agents: definitions and execution patterns}
\label{subsec:agents}

AI agents are commonly defined as goal-directed systems that generate sequences of actions while maintaining internal state that influences future decisions \cite{bousetouane2025agentic,park2023generative,bousetouane2026aiagents}. Many deployed designs follow a loop that interleaves reasoning and acting, where the model produces intermediate reasoning and then invokes tools or actions, updating state from outcomes \cite{bousetouane2025physical}. Subsequent work extends this pattern with planning modules, tool selection and tool learning, and structured execution graphs that make routing and retries explicit \cite{liu2023plan,schick2024toolformer,wu2024agentgraph}. These ideas are reflected in widely used agent frameworks that provide programmable control layers for multi-step workflows.

System-level patterns also recur across implementations. Single-agent designs emphasize sequential reasoning and tool use \cite{yao2022react}. Multi-agent designs decompose tasks across roles and coordinate through message passing \cite{li2023camel,wu2023autogen}. Human-in-the-loop designs incorporate feedback or preference signals to guide behavior \cite{bai2022rlhf} \cite{bousetouane2025agentic}. Domain-scoped agents restrict tools and objectives to improve reliability in operational workflows  \cite{chen2024vertical}. Across these variants, a shared practical challenge remains: the agent must preserve task state across turns, even as requirements evolve and distractions appear.

\subsection{Memory and state as a limiting factor in multi-turn agents}
\label{subsec:memory}

Most deployed agents represent state implicitly through text. Transcript replay preserves information by appending past turns, but prompt length grows with turn count, which increases cost and latency and reduces selectivity \cite{press2023long,liu2024lost}. Empirical studies on long context behavior also report that models can under-attend to relevant early tokens and over-attend to recent or noisy information, which contributes to drift as interactions extend \cite{liu2024lost}. These effects are especially visible in tasks where constraints and output formats must remain stable across turns.

Retrieval-based memory reduces prompt growth by storing past interactions or documents externally and retrieving relevant artifacts per turn \cite{lewis2020rag,karpukhin2020dpr}. While effective for knowledge access, retrieval optimizes for relevance signals that may not align with control relevance. Retrieved text can be locally similar yet inconsistent with the active constraints, and retrieval can surface outdated or injected artifacts, which destabilizes multi-turn control \cite{asai2023selfrag}. Memory orchestration systems introduce policies for moving information between short-term and long-term storage \cite{packer2023memgpt}, and reflection-based methods store distilled lessons or critiques \cite{shinn2023reflexion}. These approaches improve efficiency and learning, but they still treat text artifacts as the primary carrier of state, without enforcing a structured update rule for what should persist as committed state versus what should remain episodic.

This motivates a central question for agent design. Are multi-turn failures primarily a capacity problem, or do they stem from inadequate control over what becomes persistent memory?

\subsection{Cognitive compression as a principle for memory control}
\label{subsec:compression}

Cognitive science suggests that sustained multi-step reasoning is enabled by compact internal representations rather than verbatim replay. Human working memory is capacity-limited and relies on selection and executive control to maintain task-relevant state \cite{baddeley1974working,cowan2001magical}. The episodic buffer hypothesis emphasizes temporary binding into structured representations that support reasoning under constraint \cite{baddeley2000episodicbuffer}. Complementary Learning Systems theory distinguishes fast episodic acquisition from slower semantic consolidation, which reduces interference while preserving adaptability \cite{mcclelland1995cls}. Predictive coding frameworks similarly model cognition as maintaining compact internal hypotheses that are updated by prediction error \cite{rao1999predictive,friston2010free}. Recent accounts further formalize forgetting and distortion as forms of semantic compression under resource constraints, where memory is optimized to preserve decision-relevant structure rather than a faithful record of history \cite{nagy2020semanticcompression}.

These perspectives converge on a shared engineering principle. Reliable multi-turn behavior benefits from continuously distilling experience into a bounded, decision-relevant internal state rather than accumulating raw history. In the next section, we translate this principle into a memory control mechanism through the \emph{Agent Cognitive Compressor (ACC)}, which treats agent memory as an explicit compression and commitment process designed to preserve invariants while resisting drift.

\section{Agent Cognitive Compressor (ACC)}
\label{sec:acc}

The previous section established that long horizon agent failures such as drift, constraint erosion, and hallucination compounding are driven primarily by uncontrolled memory growth rather than limitations in model expressivity. Transcript replay expands context approximately linearly with interaction length, increasing non essential tokens that compete for attention and amplifying early errors through continual re exposure. Retrieval based memory reduces prompt growth, but it optimizes for semantic similarity rather than decision relevance, allowing recalled artifacts to perturb goals, constraints, and intermediate assumptions in unstable ways.

Crucially, both paradigms treat memory as accumulated text and provide no principled write path for internal state. They do not specify what should persist, what should be revised, and what should be discarded. As horizons increase, this absence of memory governance leads to drift, loss of task invariants, and hallucinations induced by irrelevant, stale, or inconsistently recalled context.

We argue, grounded in the prior work reviewed in Section~\ref{sec:background}, that long horizon reliability requires explicit memory control: a compact, structured set of decision critical variables, including goals, constraints, entities, and relations, updated selectively as new evidence arrives. This requirement is bio inspired. Humans sustain coherence by maintaining a bounded working state that is continuously regulated by salience, constraints, and expected utility.

The motivating question is: \emph{Can we equip agents with a bio inspired cognitive memory, a bounded working state that keeps them focused while reducing context growth, drift, and hallucination compounding?}

We address this question by introducing the \emph{Agent Cognitive Compressor (ACC)}, a dedicated memory controller that governs how an agent’s internal cognitive state evolves across interaction turns.

\subsection{Agent Cognitive Compressor: Definition and Role}
\label{subsec:acc_definition}

\begin{figure}[h]
    \centering
    \includegraphics[width=0.9\textwidth]{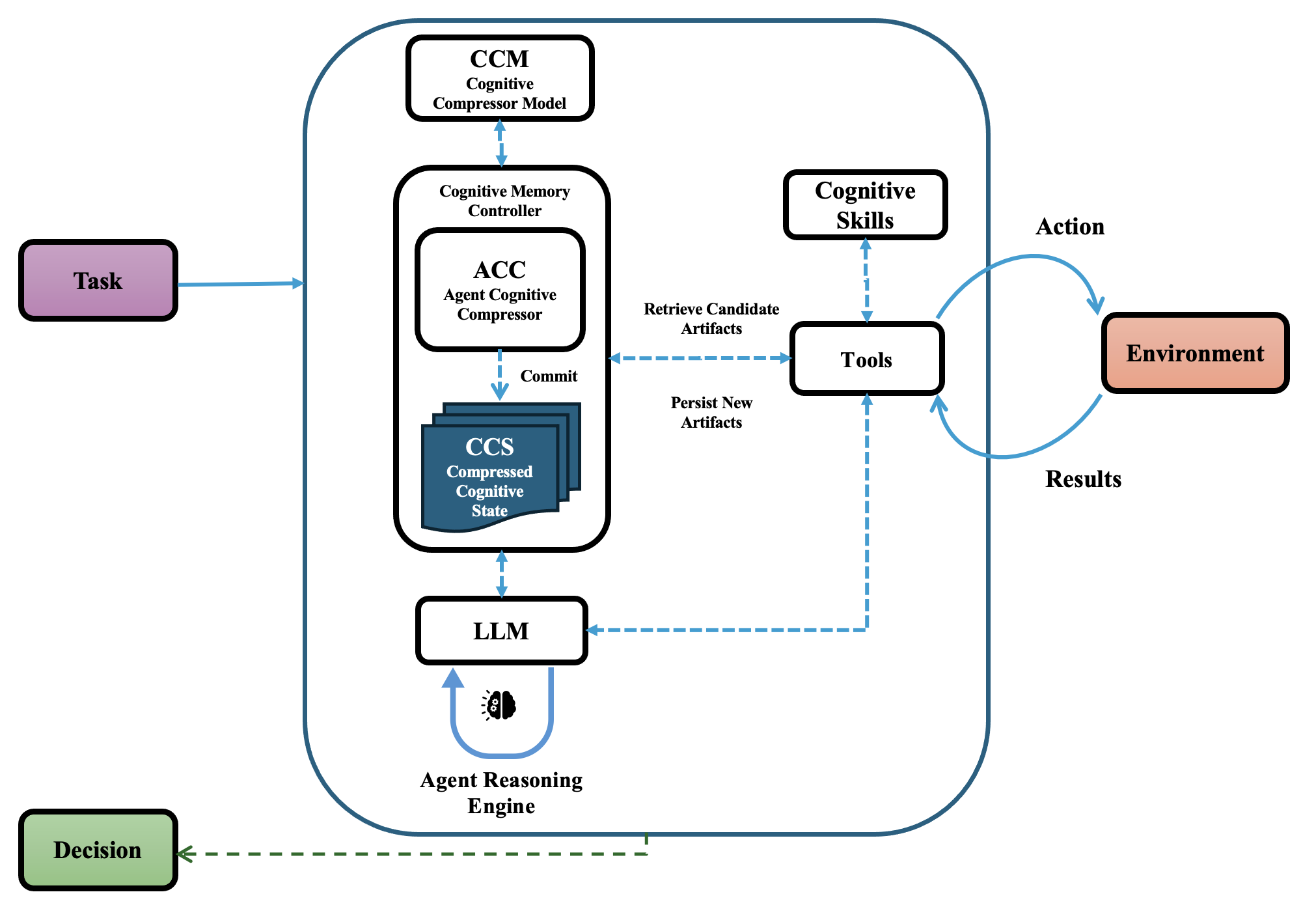}
    \caption{Agent architecture incorporating the Agent Cognitive Compressor (ACC).
ACC operates as a cognitive memory controller that constructs and commits a bounded Compressed Cognitive State (CCS) via a schema constrained Cognitive Compressor Model (CCM).
CCS is the sole persistent internal state maintained across turns and conditions downstream reasoning, tool use, and action,
while ACC remains decoupled from policy execution and environment interaction.}
    \label{fig:acc_architecture}
\end{figure}

The \emph{Agent Cognitive Compressor (ACC)} is a memory control mechanism embedded within an agent’s execution loop. ACC is not an autonomous agent and it does not implement a reasoning or acting policy. It does not perform planning, tool selection, or action selection. Its responsibility is strictly regulatory. ACC governs how the agent’s internal cognitive state is constructed, updated, and committed across interaction turns.

As illustrated in Figure~\ref{fig:acc_architecture}, ACC sits between transient interaction signals, externally stored evidence, and the agent’s reasoning engine. Rather than accumulating raw dialogue history or injecting retrieved text directly into the reasoning prompt, ACC enforces bounded persistence by writing a compact internal state that captures only what must remain stable for coherent long horizon control. The central design commitment is that ACC writes exactly one persistent state variable and updates it through controlled replacement rather than unbounded accumulation.

ACC replaces transcript replay and retrieval only context injection with an explicit internal state called the \emph{Compressed Cognitive State (CCS)}. CCS is the sole persistent internal representation carried across turns. The reasoning engine conditions on the committed state from the previous turn together with the current user input, not on the full transcript, tool traces, or raw recalled documents. This design preserves a stable signal to noise ratio as the interaction horizon grows, while keeping long horizon commitments explicit.

A defining property of ACC is that state persistence is schema governed. CCS is not a free form summary. It is a structured cognitive state constrained by an explicit schema $\mathcal{S}_{\mathrm{CCS}}$ that specifies required fields, semantic interpretation, and allowable structure. The schema acts as a contract between compression and reasoning. It makes the write path predictable, bounded, and verifiable, and it enables deterministic parsing and validation without an additional model call.

At each turn, ACC aggregates the current turn interaction signal, the previously committed cognitive state, and a bounded set of recalled artifacts from external memory. An \emph{artifact} is any externally persisted evidence that may matter for future control, including prior user constraints, prior agent commitments, tool outputs, retrieved documents, logs, or structured facts extracted earlier. Artifacts are stored outside CCS, typically in an indexed store such as a vector database augmented with metadata and provenance. Retrieval is used to propose candidate evidence, not to update internal state.

ACC then invokes a language model referred to as the \emph{Cognitive Compressor Model (CCM)} to synthesize the next committed cognitive state under the CCS schema constraint. The CCM can be implemented in two ways. A general purpose LLM can be used with a schema conditioned prompt that constrains outputs to comply with $\mathcal{S}_{\mathrm{CCS}}$. However, for deployment we recommend a fine tuned compressor specialized model, preferably a small language model, to reduce latency and cost overhead while improving schema adherence and reducing output variance. In all cases, the CCM performs abstraction, normalization, and selective incorporation of decision critical information into CCS, and it is not responsible for generating the agent’s final response.

\paragraph{Formalization.}
Let $x_t$ denote the turn level interaction signal at turn $t$, let $\mathrm{CCS}_{t-1}$ denote the previously committed cognitive state, let $\mathcal{M}$ denote an external artifact store, and let $\mathcal{S}_{\mathrm{CCS}}$ denote the CCS schema. ACC recalls a bounded set of candidate artifacts
\begin{equation}
A_t = \mathcal{R}_{\mathrm{ACC}}\!\left(x_t, \mathrm{CCS}_{t-1}; \mathcal{M}\right),
\end{equation}
filters recalled artifacts using a decision qualification gate
\begin{equation}
A_t^{+} = \left\{\, a \in A_t \;\middle|\; \mathcal{Q}\!\left(a, \mathrm{CCS}_{t-1}, x_t\right) = 1 \,\right\},
\end{equation}
and computes the next committed cognitive state under the schema constraint
\begin{equation}
\mathrm{CCS}_t =
\mathcal{C}_{\theta}\!\left(x_t, \mathrm{CCS}_{t-1}, A_t^{+}; \mathcal{S}_{\mathrm{CCS}}\right),
\end{equation}
where $\mathcal{C}_{\theta}$ denotes the schema constrained compression operator implemented by the CCM.

\paragraph{ACC flow (per turn).}
Given the formulation above, the ACC update proceeds as follows.
\begin{itemize}
    \item \textbf{Observe} the current interaction signal $x_t$ and read the previously committed state $\mathrm{CCS}_{t-1}$.
    \item \textbf{Recall} a bounded candidate set $A_t$ from the external store $\mathcal{M}$ conditioned on $(x_t,\mathrm{CCS}_{t-1})$.
    \item \textbf{Qualify} recalled items using $\mathcal{Q}$ to produce $A_t^{+}$, retaining only decision relevant evidence.
    \item \textbf{Commit} the next state by running the CCM with schema constraint $\mathcal{S}_{\mathrm{CCS}}$ to produce $\mathrm{CCS}_t$.
    \item \textbf{Replace} the prior state with the committed state, so $\mathrm{CCS}_t$ fully replaces $\mathrm{CCS}_{t-1}$.
\end{itemize}

The updated state $\mathrm{CCS}_t$ fully replaces $\mathrm{CCS}_{t-1}$ and becomes the sole persistent memory used by the agent in subsequent turns. Internal memory does not grow through accumulation. It evolves through controlled state transitions. This replacement semantics distinguishes ACC from replay based and retrieval based memory designs, and it makes the write path explicit and auditable.

\begin{figure}[t]
    \centering
    \includegraphics[width=\linewidth]{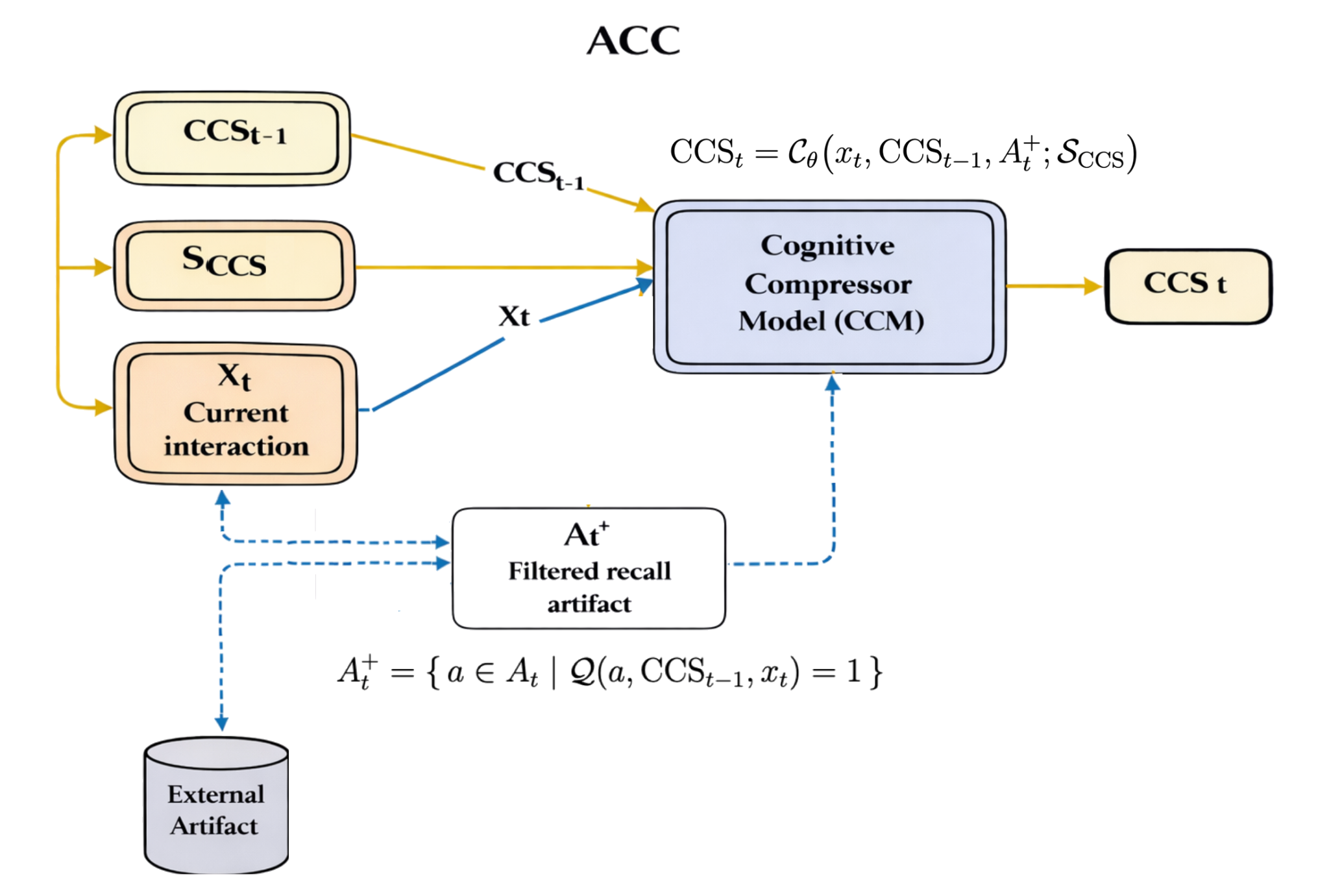}
    \caption{ACC state commitment mechanism for producing the next Compressed Cognitive State $\mathrm{CCS}_t$ under the schema constraint $\mathcal{S}_{\mathrm{CCS}}$, using the current interaction $x_t$, the previously committed state $\mathrm{CCS}_{t-1}$, and the qualified recalled set $A_t^{+}$.}
    \label{fig:acc_ccs_update_flow}
\end{figure}

Finally, ACC enforces a strict separation between artifact recall and state commitment. Retrieved artifacts are treated as external evidence and cannot directly modify the cognitive state unless explicitly incorporated by the CCM under the CCS schema constraint. This separation limits uncontrolled state mutation, reduces sensitivity to retrieval noise, and improves long horizon stability. Figure~\ref{fig:acc_ccs_update_flow} provides a detailed view of the ACC state commitment mechanism used to produce $\mathrm{CCS}_t$.

The following subsection defines the Compressed Cognitive State (CCS) itself and details its schema, cognitive motivation, and role as the minimal internal representation required to sustain coherent reasoning over extended interactions.

\subsection{Compressed Cognitive State (CCS)}
\label{subsec:ccs}

The \emph{Compressed Cognitive State (CCS)} is the bounded internal representation maintained by ACC across interaction turns. CCS is neither a transcript summary nor a memory cache. Instead, it is a control oriented cognitive state designed to preserve decision critical information while explicitly discarding irrelevant, redundant, or low utility detail.

CCS is bio inspired. Its design is motivated by how humans maintain task relevant mental state during extended problem solving, not by replaying experience verbatim, but by retaining a compact internal representation of what changed, what matters, what constraints apply, and what goal is being pursued. Cognitive science and neuroscience literature consistently show that human working cognition relies on abstraction, salience, and relational structure rather than raw episodic recall. CCS mirrors this principle by encoding only the minimal set of variables required to sustain coherent reasoning over time \cite{baddeley2012, miller1956, norman1975, cowan2001, kahneman2011}. In ACC, this cognitive principle is operationalized as typed, bounded fields that make the agent’s commitments explicit rather than latent in a growing transcript.

Accordingly, CCS is defined as a schema with explicit, typed components that correspond to functional elements of human cognition. This schema is not merely descriptive documentation. It is actively enforced during execution. At each turn, ACC prompts the Cognitive Compressor Model (CCM) to emit a CCS instance conforming to the schema $\mathcal{S}_{\mathrm{CCS}}$, after which the output is deterministically parsed and validated. This validation step does not require an additional model call. Enforcing schema compliance reduces output variance, prevents missing or malformed state, and makes CCS instances auditable, comparable, stable, and instrumentable across turns.

Each component of the CCS schema serves a distinct functional role analogous to a cognitive function in human reasoning. Together, they form a minimal decomposition of the variables required to maintain coherent long horizon control:

\begin{itemize}
    \item \textbf{Episodic trace} captures what changed in the current turn, analogous to short term event updating in human working memory.
    \item \textbf{Semantic gist} encodes the dominant intent or topic, reflecting how humans abstract meaning beyond surface form.
    \item \textbf{Focal entities (typed)} represent canonicalized objects or actors, supporting stable reference resolution over long horizons.
    \item \textbf{Relational map} encodes causal and temporal dependencies, mirroring human causal and temporal reasoning.
    \item \textbf{Goal orientation} represents the persistent objective guiding the interaction, analogous to goal maintenance in executive control.
    \item \textbf{Constraints} encode task, policy, or safety rules that remain invariant, similar to rule based inhibition in human cognition.
    \item \textbf{Predictive cue} captures the expected next cognitive operation, resembling anticipatory planning in problem solving.
    \item \textbf{Uncertainty signal} explicitly represents unresolved or low confidence state, discouraging overconfident inference.
    \item \textbf{Retrieved artifacts} retain references to external evidence, analogous to consulting notes or tools without internalizing raw content.
\end{itemize}

Figure~\ref{fig:ccs_toon_example} illustrates a CCS instance for an IT operations scenario, serialized using a lightweight TOON style token oriented representation. TOON here denotes a compact token oriented serialization designed to minimize entropy while preserving legibility. Compared to transcript replay, this representation is intentionally decision oriented. The encoding exposes CCS as a structured cognitive state composed of typed entities, persistent goals, explicit constraints, and stable relational predicates, while external evidence is retained as referenced artifacts with provenance rather than injected text. This design makes the agent’s internal state interpretable, bounded, and directly usable for downstream reasoning and control.

\begin{figure}
\centering
\caption{Example Compressed Cognitive State (CCS) for an IT operations task, serialized using a TOON style token oriented representation.}
\label{fig:ccs_toon_example}

\footnotesize
\fbox{
\begin{minipage}{0.96\linewidth}
\setlength{\parindent}{0pt}
\setlength{\parskip}{0.35em}

\textbf{user\_query:}\\
We see intermittent 502 errors in Nginx after enabling HTTP/2. I cannot restart during business hours. Give a safe mitigation and the next checks.

\textbf{ccs\_t:}

\textbf{episodic\_trace:}\\
observed(502\_spikes after(enable(http2)))\\
logged(nginx\_error upstream\_closed\_early)\\
constraint(no\_restart during(business\_hours))

\textbf{semantic\_gist:}\\
mitigate(502) \& diagnose(upstream\_instability)

\textbf{focal\_entities:}\\
host(vm\_ubuntu22\_04)\;
service(nginx)\;
service(node\_upstream)\;
feature(http2)\;
signal(error\_502)

\textbf{relational\_map:}\\
timing(502\_spikes after(http2\_enable))\\
possible(upstream\_timeout 502)\\
possible(upstream\_connection\_close 502)

\textbf{goal\_orientation:}\\
reduce(502\_rate within(10min)) \& preserve(service\_availability)

\textbf{constraints:}\\
no\_restart(nginx)\;
reload\_allowed(nginx)\;
safe\_change(minimal)\;
avoid(speculation)

\textbf{predictive\_cue:}\\
check(upstream\_latency)\; check(node\_memory\_growth)\; validate(nginx\_timeouts)

\textbf{uncertainty\_signal:}\\
level(medium)\; gaps(root\_cause not\_confirmed)

\textbf{retrieved\_artifacts:}\\
nginx\_error\_snippet(upstream\_prematurely\_closed)\\
recent\_change(enable\_http2)\;
constraint\_note(no\_restart)

\end{minipage}
}
\end{figure}

Crucially, CCS is not static. It is continuously constructed, evaluated, and selectively updated at each interaction turn. The following section describes how CCS participates in the ACC multiturn control loop, governing recall, filtering, prediction, and state commitment over time.

\newpage
\subsection{ACC Multiturn Control Loop}
\label{subsec:acc_loop}

ACC implements schema governed memory control as a per turn state transition system over a single persistent variable, $\mathrm{CCS}_t$. At turn $t$, the agent policy conditions on the current user input together with the previously committed state. The completed turn is then treated as the interaction signal $x_t$ for ACC, which performs recall, qualification, and schema constrained commitment to produce the next state. Algorithm~\ref{alg:acc_loop} summarizes the multi-turn ACC update process. At each turn, ACC (i) recalls a bounded set of candidate artifacts from $\mathcal{M}$, (ii) applies a qualification gate to retain only decision-relevant evidence, and (iii) commits the next bounded state $\mathrm{CCS}_t$ under the schema constraint $\mathcal{S}_{\mathrm{CCS}}$.

\begin{algorithm}[t]
\caption{ACC multiturn cognitive state commitment and evidence persistence}
\label{alg:acc_loop}
\begin{algorithmic}[1]
\Require Initial committed state $\mathrm{CCS}_0$, external artifact store $\mathcal{M}$, CCS schema $\mathcal{S}_{\mathrm{CCS}}$, horizon $T$
\State $\mathrm{CCS} \gets \mathrm{CCS}_0$
\For{$t=1$ to $T$}
    \State \textbf{Input:} observe turn interaction signal $x_t$
    \State \hspace{1.2em}{\small\emph{// $x_t$ is the turn level interaction record provided to ACC}}
    \vspace{0.25em}
    \State $A_t \gets \mathcal{R}_{\mathrm{ACC}}(x_t, \mathrm{CCS}; \mathcal{M})$
    \State \hspace{1.2em}{\small\emph{// recall a bounded candidate artifact set from $\mathcal{M}$}}
    \vspace{0.25em}
    \State $A_t^{+} \gets \{\, a \in A_t \mid \mathcal{Q}(a, \mathrm{CCS}, x_t)=1 \,\}$
    \State \hspace{1.2em}{\small\emph{// filter recalled artifacts by decision qualification}}
    \vspace{0.25em}
    \State $\mathrm{CCS}_t \gets \mathcal{C}_{\theta}\big(x_t, \mathrm{CCS}, A_t^{+}; \mathcal{S}_{\mathrm{CCS}}\big)$
    \State \hspace{1.2em}{\small\emph{// commit the next state via the schema constrained CCM}}
    \vspace{0.25em}
    \State \textbf{Commit:} $\mathrm{CCS} \gets \mathrm{CCS}_t$
    \State \hspace{1.2em}{\small\emph{// $\mathrm{CCS}_t$ fully replaces the previously committed state}}
    \vspace{0.25em}
    \State \textbf{Agent execution:} $\mathrm{decision}_t \gets \mathrm{Agent}(\mathrm{CCS}_t,\mathrm{role},\mathrm{tools})$
    \State \hspace{1.2em}{\small\emph{// $\mathrm{decision}_t$ includes the turn output and any tool actions}}
    \vspace{0.25em}
    \State \textbf{Evidence persistence:} $\mathcal{M} \gets \mathcal{M} \cup \mathrm{Store}(x_t, \mathrm{decision}_t)$
    \State \hspace{1.2em}{\small\emph{// store the new interaction as future evidence for recall}}
    \vspace{0.35em}
\EndFor
\end{algorithmic}
\end{algorithm}

This loop makes the control semantics explicit. The only persistent internal memory is the committed state, updated once per turn through a schema constrained commitment operation. Recall produces candidate evidence, qualification restricts which evidence may be considered, and commitment determines what is incorporated into $\mathrm{CCS}_t$. The result is a bounded, inspectable, and repeatable update mechanism that supports stable long horizon behavior by construction.

\begin{itemize}
    \item \textbf{Bounded persistence.} The internal state remains constant in size across turns because only $\mathrm{CCS}_t$ persists.
    \item \textbf{Explicit commitment.} State changes occur only via $\mathcal{C}_{\theta}(\cdot)$ under $\mathcal{S}_{\mathrm{CCS}}$, which makes updates auditable and comparable across turns.
    \item \textbf{Robust recall interface.} Retrieval proposes $A_t$ but only qualified evidence $A_t^{+}$ is eligible to influence commitment.
    \item \textbf{Stable conditioning.} The policy $\pi$ receives a fixed format state rather than a growing transcript, reducing sensitivity to horizon length.
    \item \textbf{Drift and hallucination suppression.} Commitments are refreshed through structured state transition, while irrelevant or stale context is filtered before it can affect the next state.
\end{itemize}

\section{ACC in Agent Architecture Patterns}
\label{subsec:acc_in_architectures}

\begin{figure*}[t]
    \centering
    \begin{subfigure}[t]{0.48\textwidth}
        \centering
        \includegraphics[width=\textwidth]{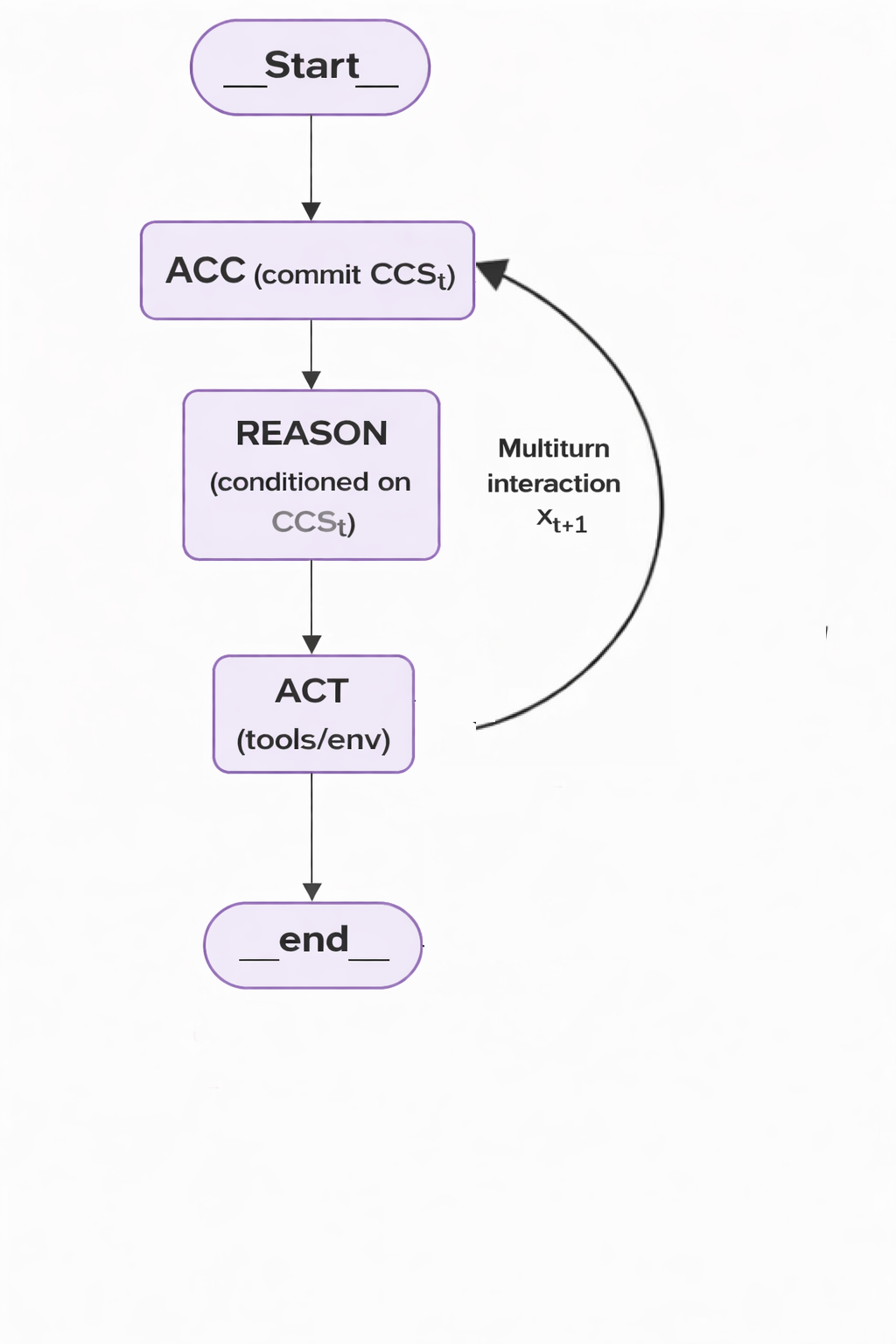}
        \caption{ReAct multi-turn loop with ACC committing $\mathrm{CCS}_t$ prior to \texttt{REASON}, and a multi-turn return from \texttt{ACT} to \texttt{ACC}.}
        \label{fig:acc_react_flow}
    \end{subfigure}
    \hfill
    \begin{subfigure}[t]{0.48\textwidth}
        \centering
        \includegraphics[width=\textwidth]{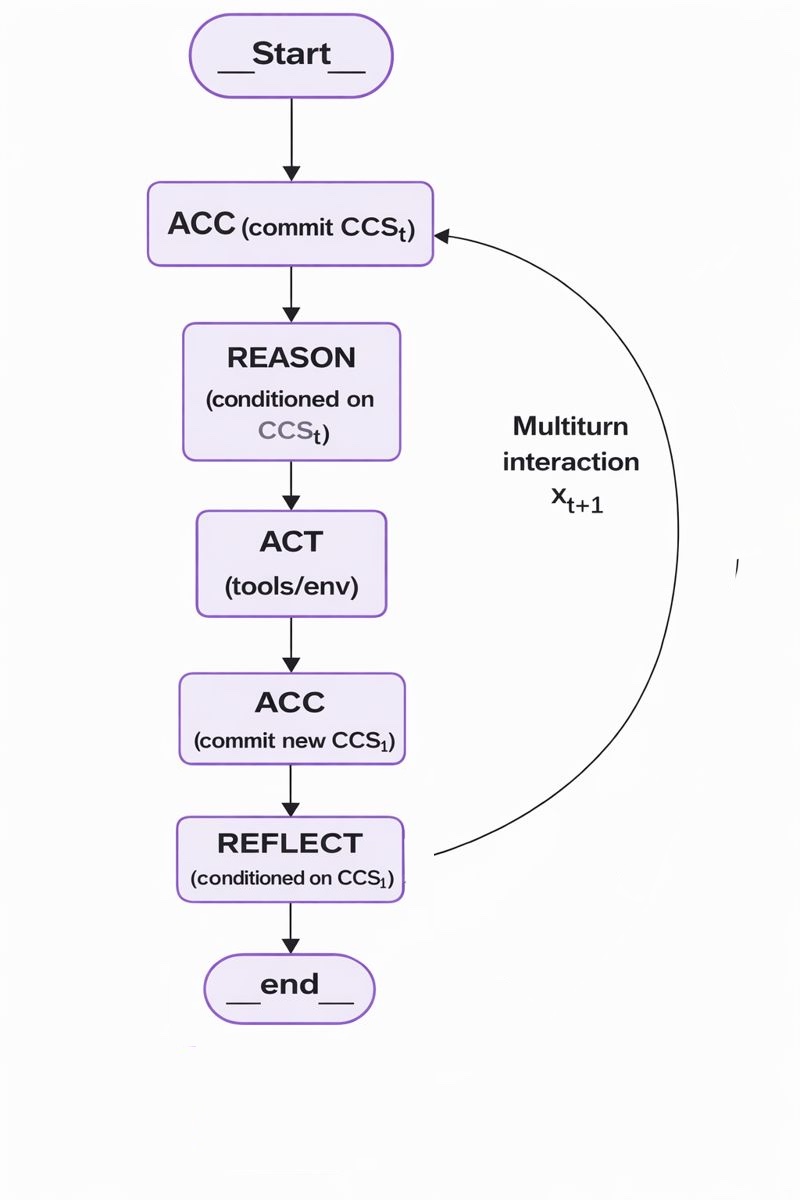}
        \caption{Planning agent loop: \texttt{ACC} $\rightarrow$ \texttt{REASON (plan)} $\rightarrow$ \texttt{ACT} $\rightarrow$ \texttt{ACC} $\rightarrow$ \texttt{REFLECT} (stop or repeat).}
        \label{fig:acc_planning_flow}
    \end{subfigure}
    \caption{ACC integration patterns in agent architectures: ReAct multi-turn tool execution (left) and plan execute reflect control (right).}
    \label{fig:acc_two_patterns}
\end{figure*}

ACC integrates into common agent control loops by acting as a memory control mechanism that commits a bounded $\mathrm{CCS}_t$ at each turn. We describe two patterns that cover a large portion of enterprise agent deployments: a ReAct multi-turn tool agent and a planning agent with an explicit plan, execute, reflect cycle. Figure~\ref{fig:acc_two_patterns} summarizes both flows.

\subsection{Pattern 1: Multi-turn ReAct Agent}
\label{subsubsec:acc_react}

In ReAct-style agents, execution alternates between reasoning and tool actions. ACC is invoked once per turn to commit $\mathrm{CCS}_t$, which then conditions the next reasoning step (Fig.~\ref{fig:acc_react_flow}). This makes long-run consistency depend on a bounded state rather than on replaying a growing transcript or injecting retrieved text into the prompt.

\newpage
In this pattern, the turn flow is explicit.
\begin{enumerate}
    \item ACC commits $\mathrm{CCS}_t$ from the current interaction signal $x_t$, the prior state $\mathrm{CCS}_{t-1}$, and qualified evidence.
    \item The reasoning step uses $(x_t, \mathrm{CCS}_t)$ to decide the next diagnostic or action.
    \item The agent performs \texttt{ACT}, producing tool outputs and intermediate observations.
    \item The next turn returns to ACC, which updates the state by replacing $\mathrm{CCS}_t$ with $\mathrm{CCS}_{t+1}$.
\end{enumerate}

\paragraph{Example: healthcare operations incident triage (multi-turn).}
Consider an agent supporting an analytics team during an intermittent ingestion failure in a clinical data pipeline. The conversation spans many turns as the user provides new evidence and requests safe mitigations under operational constraints. Tool outputs are verbose and repetitive.

ACC preserves a compact set of invariants while evidence grows externally.
\begin{itemize}
    \item \textbf{Constraints that must hold.} For example, no disruptive restarts during clinic hours and only read-only diagnostics until the maintenance window.
    \item \textbf{Verified entities and scope.} For example, pipeline name, environment, affected job IDs, and time window.
    \item \textbf{Progress and commitments.} For example, checks already executed, active hypothesis, and the next best diagnostic.
\end{itemize}
Without ACC, transcript replay amplifies noise and early guesses, and retrieval-only memory can surface stale runbooks that perturb control. With ACC, the agent stays anchored to committed constraints and verified state, which reduces repetition and drift across turns.

\subsection{Pattern 2: Multi-turn  Planning Agent }
\label{subsubsec:acc_planning}

Planning agents make the plan explicit and include a reflection step that decides whether to stop, revise, or continue. ACC strengthens this loop by maintaining bounded plan commitments and validated facts across iterations (Fig.~\ref{fig:acc_planning_flow}). The planning flow is typically:
\begin{enumerate}
    \item ACC commits $\mathrm{CCS}_t$ capturing the current goal, constraints, and validated state.
    \item The reasoning step proposes or updates a plan conditioned on $\mathrm{CCS}_t$.
    \item The agent executes one or more steps, often using web search or enterprise search tools.
    \item ACC updates the state to $\mathrm{CCS}_{t+1}$ by committing only validated outcomes and stable commitments.
    \item Reflection uses $\mathrm{CCS}_{t+1}$ to decide whether to stop or repeat with targeted revisions.
\end{enumerate}

\paragraph{Example: web-grounded remediation planning.}
Consider a planning agent drafting a remediation plan for a newly disclosed software vulnerability using web search. Search results are noisy: some pages are outdated, speculative, or biased.

ACC prevents unverified content from becoming persistent state by committing only:
\begin{itemize}
    \item \textbf{Validated facts with provenance.} For example, affected versions and official mitigation steps from authoritative sources.
    \item \textbf{Hard constraints.} For example, maintenance window, approval gates, and communication policy.
    \item \textbf{Plan commitments and progress.} For example, selected option, open questions, and the next evidence needed.
\end{itemize}
This reduces drift under conflicting sources and makes reflection more reliable because it operates over a stable, bounded record of what is confirmed, what is constrained, and what remains uncertain.

Across both patterns, ACC enables the same core property: external content and tool outputs can scale with task complexity, while the internal state that drives decisions remains bounded, structured, and resistant to drift.

\section{Agent Judge Driven Multi-Turn Evaluation}
\label{sec:evaluation}

Evaluating AI agents is difficult because many tasks are open ended, success depends on multi-turn continuity, and errors compound as interactions extend. Recent work shows that LLM-based judging can approximate human preference for open ended answers and enables scalable comparisons without gold labels \cite{zheng2023judging,liu2023geval}. Agent benchmarks further highlight that failures in multi-step settings are often driven by instruction-following breakdowns and multi-turn consistency issues rather than single-turn fluency \cite{liu2023agentbench}. At the same time, surveys emphasize that agent evaluation remains fragmented, with no single standard that jointly measures task outcomes and memory reliability in multi-turn settings \cite{guan2025multiturnsurvey,yehudai2025surveyagents}.

This gap is central to ACC. ACC targets stability by controlling what persists in memory, so evaluation must quantify not only response quality, but also memory-driven failures such as hallucination carryover and drift from established constraints. We introduce an evaluation framework for multi-turn agents that couples response-driven outcome scoring with memory-driven auditing of hallucination and drift, under a single judge-controlled live testing process.

\subsection{Evaluation setup and judge agent}
We compare three different agents under identical multi-turn scenarios: a baseline agent that replays transcript history, a retrieval agent that retrieves context from its own isolated store, and the proposed ACC agent that carries a bounded Compressed Cognitive State. We evaluate across multiple domains to avoid conclusions that are specific to one task family. Our scenarios cover IT operations, cybersecurity response, healthcare operations, and finance workflows.

The evaluation is driven by an agent judge, not a static grader prompt. For each turn, the judge issues a live query, then all three agents answer. In our implementation we execute agents in a fixed live order, baseline then retrieval then ACC agent, and then call the judge to score the three answers for that turn. The judge maintains a canonical memory updated from user queries and prior canonical memory only, and uses that canonical memory as the reference for both scoring and memory auditing.

\subsection{Response-driven outcome evaluation}
At turn $t$, the judge query is $x_t$ and agent $a$ returns answer $y_t^{a}$. The judge assigns outcome scores on a 0 to 10 scale for a small set of task oriented metrics that remain stable across domains.
\begin{enumerate}
  \item Relevance, alignment to the request and constraints
  \item Answer quality, usefulness and correctness for the task
  \item Instruction following, compliance with explicit format and limits
  \item Coherence, clarity and internal consistency
\end{enumerate}

We denote the score for metric $m$ as $s_{t,m}^{a} \in [0,10]$. For each metric we report the mean score across turns
\[
\bar{s}_{m}^{a} = \frac{1}{T}\sum_{t=1}^{T} s_{t,m}^{a},
\]
and we report variability across turns to capture stability under extended multi-turn conditions. To reduce judge bias, we blind agent identities during scoring, randomize presentation order, and bound visible context length. We follow established recommendations to control position and verbosity effects and to log the full judge configuration for reproducibility \cite{zheng2023judging,gu2024llmasjudge,guerdan2025validating}.

\subsection{Memory-driven evaluation: hallucination and drift}
Outcome scores alone can miss memory failures that emerge over time. We therefore add two memory-driven evaluations grounded in the judge canonical memory.

\paragraph{Hallucination audit.}
For each answer $y_t^{a}$, the judge extracts a set of atomic claims and labels each claim as supported or unsupported using evidence restricted to the current user query and the judge-maintained canonical memory. Let $S_t^{a}$ and $U_t^{a}$ denote the number of supported and unsupported claims at turn $t$, respectively. We compute a turn-level hallucination rate as
\[
H_t^{a} = \frac{U_t^{a}}{\max(1, S_t^{a} + U_t^{a})}.
\]
Lower values of $H_t^{a}$ indicate better grounding. We report the average hallucination rate across turns as
\[
H_u^{a} = \frac{1}{T}\sum_{t=1}^{T} H_t^{a},
\]
and plot $H_t^{a}$ over time to identify hallucination bursts, which commonly occur following misleading or adversarial turns.

\paragraph{Drift audit.}
Drift is measured as deviation from previously established constraints and required outputs. At each turn, the judge reads the current query and the prior canonical memory, extracts the active requirements and constraints, and checks whether $y_t^{a}$ violates them or omits required elements. Let $V_t^{a}$ denote the number of constraint violations and $O_t^{a}$ the number of omitted required elements at turn $t$, and let $\mathcal{K}_t$ denote the set of active constraints at that turn. We compute a normalized drift rate as
\[
D_t^{a} = \frac{V_t^{a} + O_t^{a}}{\max(1, |\mathcal{K}_t|)}.
\]
Lower values of $D_t^{a}$ indicate stronger multi-turn consistency. Drift is undefined at the first turn due to the absence of a prior state, so we exclude $t=1$ and report the average drift rate as
\[
D_r^{a} = \frac{1}{T-1}\sum_{t=2}^{T} D_t^{a}.
\]

\paragraph{Memory footprint.}
Because ACC is a memory control mechanism, we also log memory growth. We report the per turn token size of baseline transcript memory, retrieval context, and ACC state. Let $M_t^{a}$ denote the measured memory size at turn $t$ for agent $a$. We plot $M_t^{a}$ across turns and report summary statistics such as $\bar{M}^{a}$ and $M_T^{a}$. This links memory consumption to hallucination and drift behavior in multi-turn scenarios.

% (If not already in your preamble)
% \usepackage{subcaption}

\subsection{Evaluation data and scenarios}
\label{subsec:eval_data}

The evaluation data consists of judge-generated questions grounded in curated scenarios. We launched 600 live evaluations for a total of 30{,}000 turns across all data. Each live evaluation runs for 50 turns to evaluate extended multi-turn behavior, including token consumption, goal sustaining, constraint retention, and robustness under noisy and adversarial updates. We evaluate across three domains:  IT operations, cybersecurity response, and healthcare workflows

Scenario design mixes standard task progression with stress turns that probe common multi-turn failure modes. Table~\ref{tab:data_topics} summarizes the stress topics embedded throughout the 50-turn live testing. The table serves as a dataset taxonomy that clarifies which reliability risks are covered and how the same testing surface is applied consistently across domains.

\begin{table}[t]
\centering
\small
\begin{tabular}{p{0.28\linewidth} p{0.64\linewidth}}
\hline
\textbf{Topic in the data} & \textbf{Purpose in evaluation} \\
\hline
Constraints and environment setup & Establish baseline requirements and operational limits early \\
Evidence updates and recalibration & Force hypothesis updates without rewriting prior facts \\
Strict instruction following & Enforce templates, length limits, and structured outputs \\
Conflicting requirements and drift traps & Test resistance to goal deviation and constraint violations \\
Bias pressure from stakeholders & Test factual tone and avoidance of motivated conclusions \\
Poisoned runbook snippets & Test safe critique and refusal of unsafe shortcuts \\
Prompt injection attempts & Test refusal to override prior constraints or policies \\
Uncertainty handling & Test separation of confirmed facts versus hypotheses \\
Operational risk tradeoffs & Test conservative decision making under pressure \\
Executive and incident communication & Test concise, accurate summaries without invented details \\
\hline
\end{tabular}
\caption{Dataset taxonomy of stress topics embedded in the judge-generated evaluation data. Each topic is instantiated with domain-specific content across IT operations, cybersecurity, healthcare, and finance.}
\label{tab:data_topics}
\end{table}

\newpage
\subsection{Evaluation results and analysis}
\label{subsec:eval_results}

We report results at two complementary levels. First, we aggregate response-driven outcome metrics across domains to quantify overall task performance under multi-turn stress (Fig.~\ref{fig:avg_outcome_metrics_domains}). Second, we report memory reliability and memory footprint behavior over 50-turn episodes, including turn-level hallucination and drift audits (Fig.~\ref{fig:audits_all_domains}) and token growth across domains (Fig.~\ref{fig:memory_all_domains}). Together, these views capture both \emph{what} the agent produces (task outcome) and \emph{how reliably} it sustains state over long horizons (hallucination, drift, and memory growth).

\subsubsection{Memory footprint across domains}
\begin{figure*}
\centering
\begin{subfigure}[t]{0.49\textwidth}
    \centering
    \includegraphics[width=\textwidth]{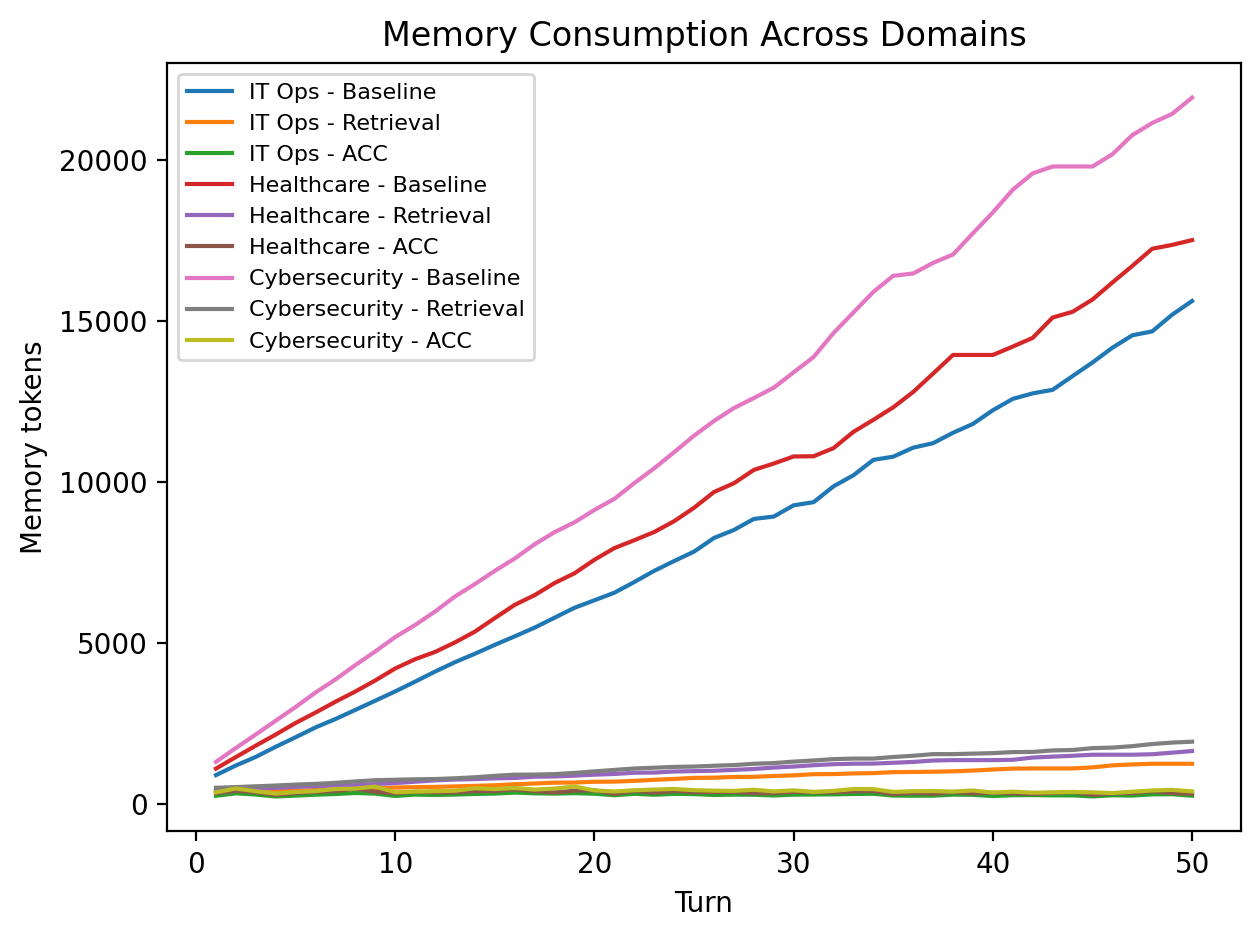}
    \caption{Per-turn memory tokens across domains.}
    \label{fig:mem_tokens_all_domains}
\end{subfigure}
\hfill
\begin{subfigure}[t]{0.49\textwidth}
    \centering
    \includegraphics[width=\textwidth]{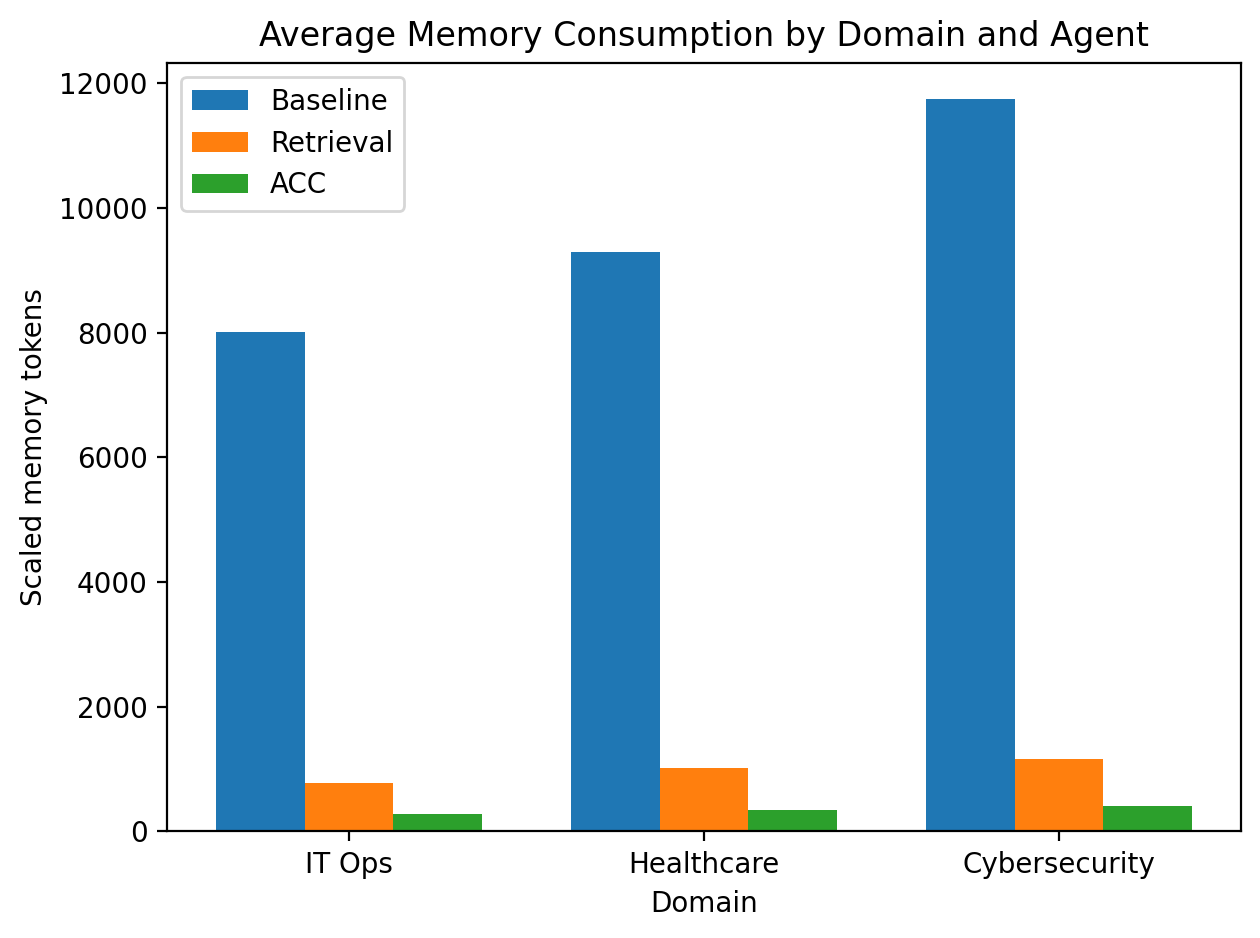}
    \caption{Average memory consumption by domain and agent.}
    \label{fig:avg_mem_tokens_by_domain_agent}
\end{subfigure}
\caption{Memory footprint across domains over 50-turn live testing. Baseline grows with turn count, while Retrieval (top-3 artifacts) and ACC remain bounded.}
\label{fig:memory_all_domains}
\end{figure*}

Figure~\ref{fig:memory_all_domains} shows a clear separation between memory mechanisms across domains. The baseline agent exhibits near-linear growth in context size over turns, increasing the amount of history that competes for attention. In contrast, both the retrieval agent and the ACC agent remain bounded across turns. For retrieval, we explicitly restrict recall to the top 5 artifacts per turn to limit drift escalation observed with larger retrieval sets. The ACC agent remains bounded by design because it carries a compact Compressed Cognitive State rather than replaying a growing transcript.

\subsubsection{Response-driven outcome metrics}
\begin{figure}
\centering
\includegraphics[width=0.95\linewidth]{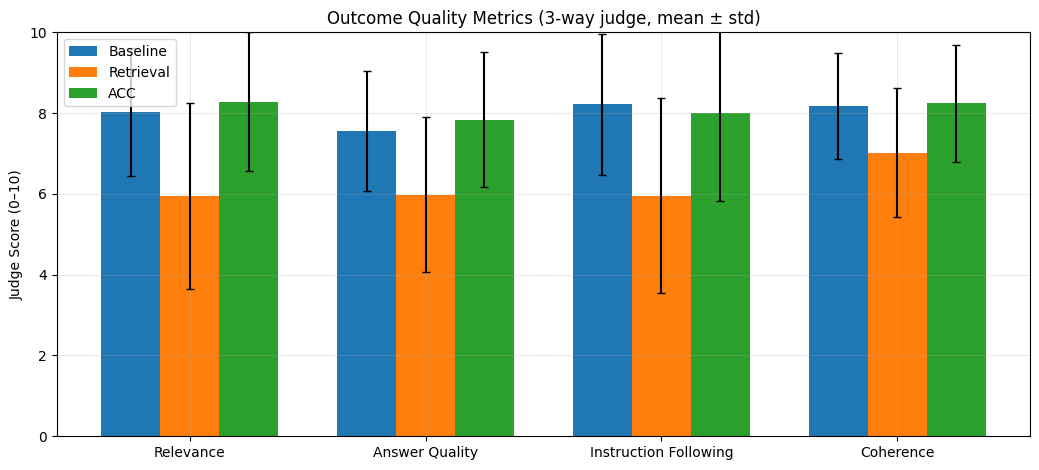}
\caption{Outcome quality metrics (three-way judge, mean $\pm$ std across turns) aggregated across domains for Relevance, Answer Quality, Instruction Following, and Coherence.}
\label{fig:avg_outcome_metrics_domains}
\end{figure}

Figure~\ref{fig:avg_outcome_metrics_domains} summarizes outcome quality metrics produced by the three-way judge. Across domains, the ACC agent achieves the strongest relevance and coherence, while remaining competitive on answer quality and instruction following. This pattern is consistent with ACC providing a stable, decision-oriented conditioning signal as horizon length increases.

\subsubsection{Hallucination rate over turns (claim audit)}
Figure~\ref{fig:audits_all_domains} (left column) reports the turn-level hallucination rate $H_t^{a}$ computed by the judge claim audit grounded in canonical memory. Across the shown domains, hallucination increases with turn count for the baseline and retrieval agents, with recurrent spikes after stress turns. The ACC agent remains near zero across most turns with only small transient spikes, indicating that unsupported claims are less likely to enter and persist in the committed cognitive state used for conditioning.

\subsubsection{Drift rate and constraint retention}
\begin{figure}
\centering
% Row 1: IT Ops
\begin{subfigure}[t]{0.49\textwidth}
    \centering
    \includegraphics[width=\textwidth]{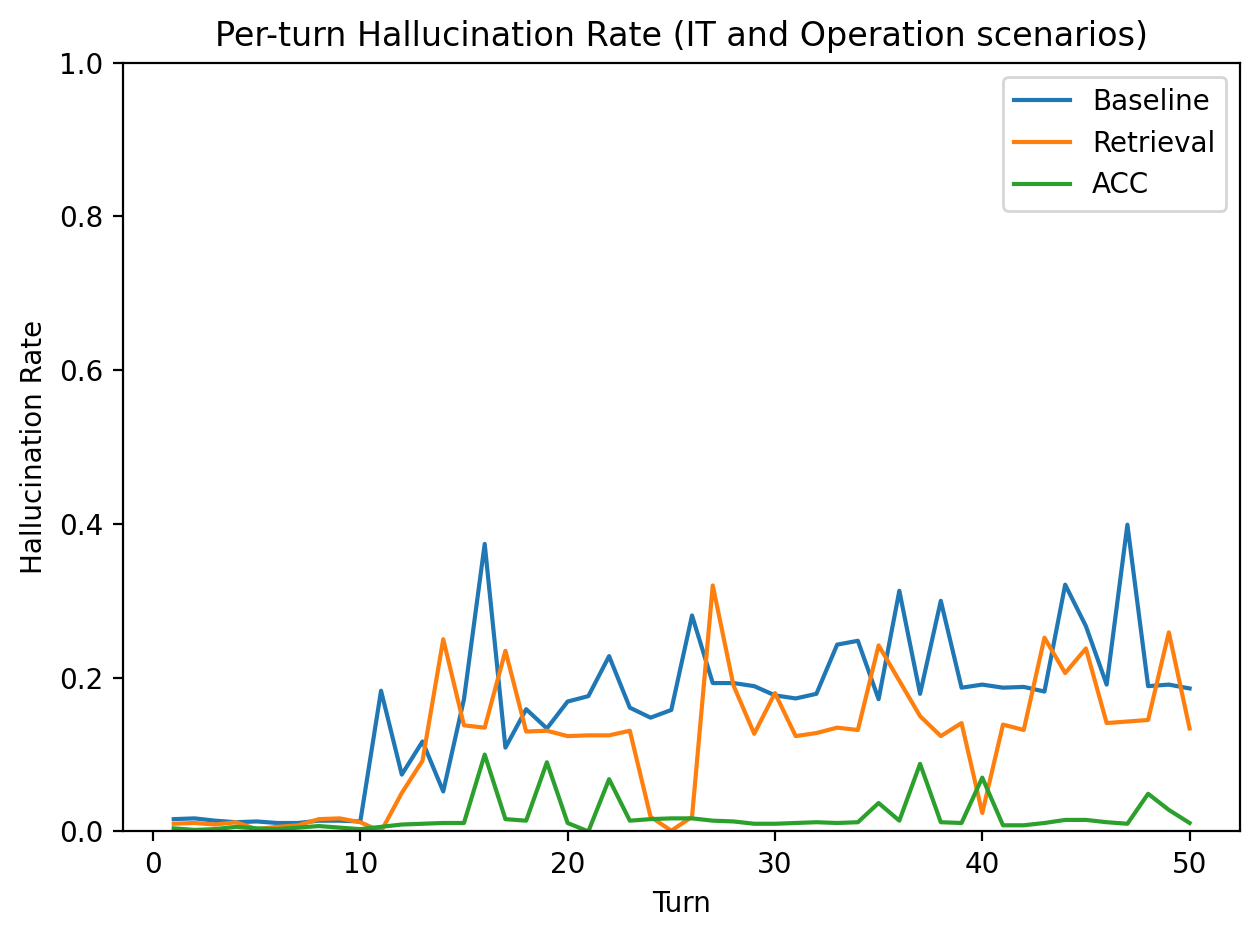}
    \caption{IT operations: hallucination rate $H_t^{a}$.}
    \label{fig:it_hallucination}
\end{subfigure}
\hfill
\begin{subfigure}[t]{0.49\textwidth}
    \centering
    \includegraphics[width=\textwidth]{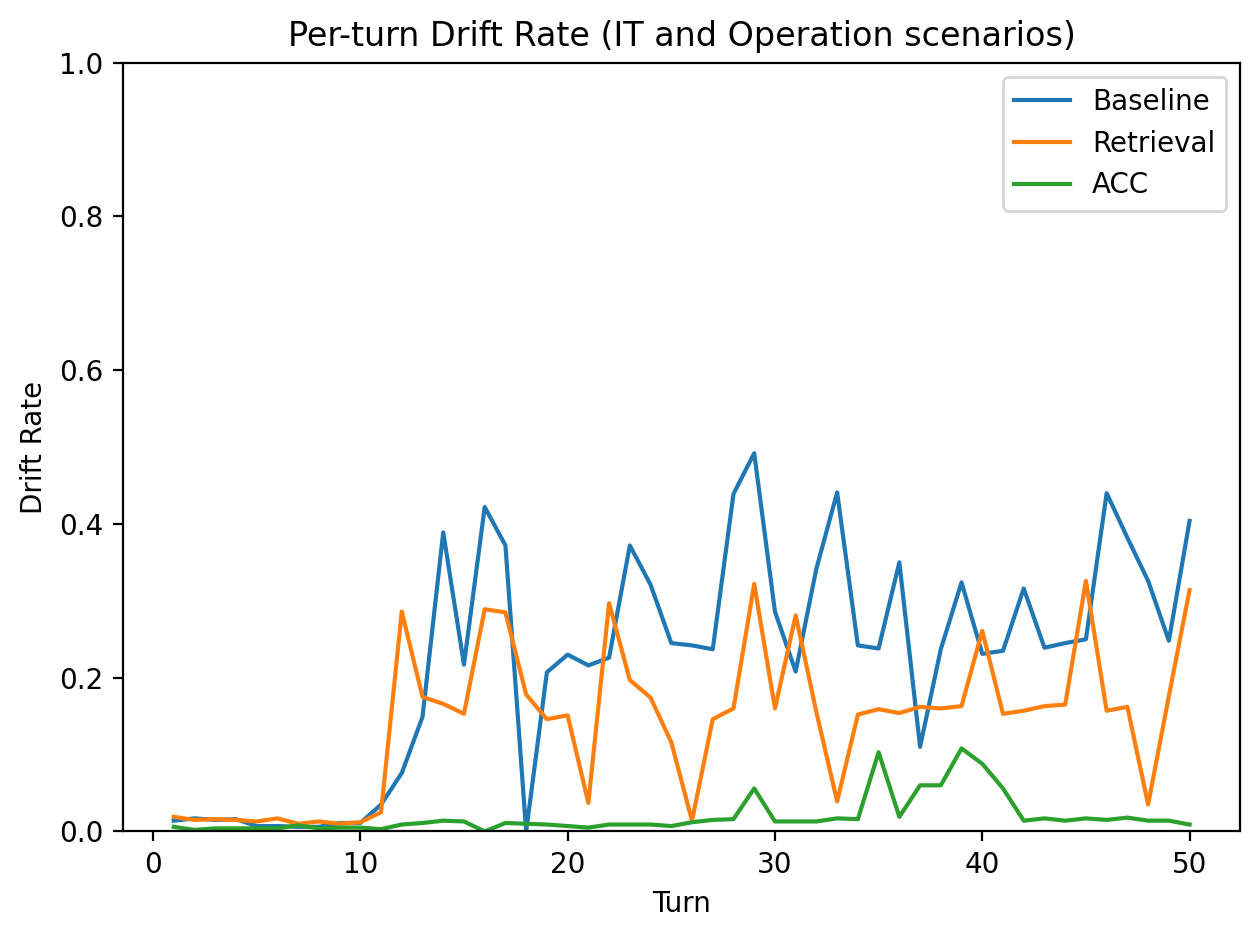}
    \caption{IT operations: drift rate $D_t^{a}$.}
    \label{fig:it_drift}
\end{subfigure}

\vspace{0.7em}
% Row 2: Healthcare
\begin{subfigure}[t]{0.49\textwidth}
    \centering
    \includegraphics[width=\textwidth]{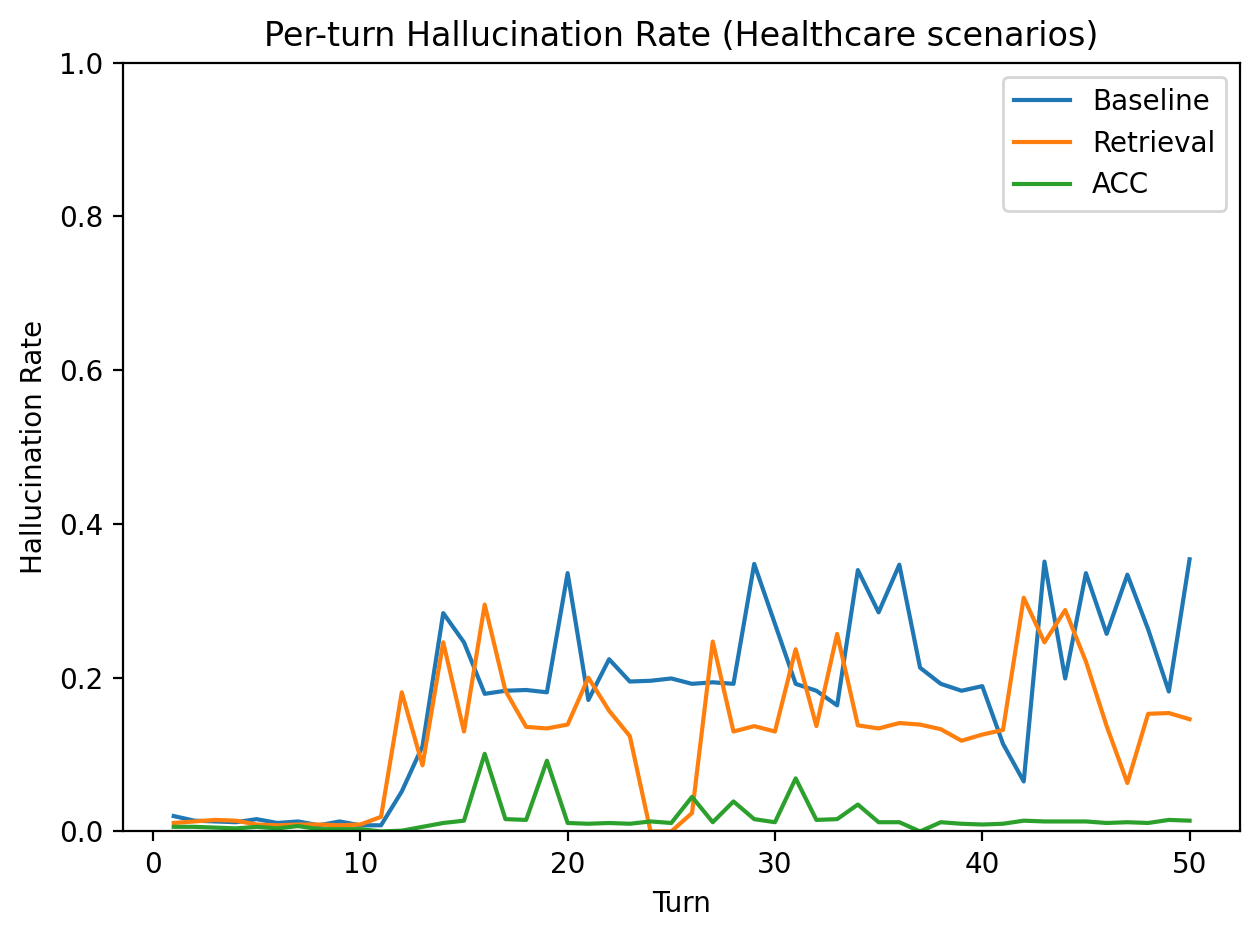}
    \caption{Healthcare: hallucination rate $H_t^{a}$.}
    \label{fig:health_hallucination}
\end{subfigure}
\hfill
\begin{subfigure}[t]{0.49\textwidth}
    \centering
    \includegraphics[width=\textwidth]{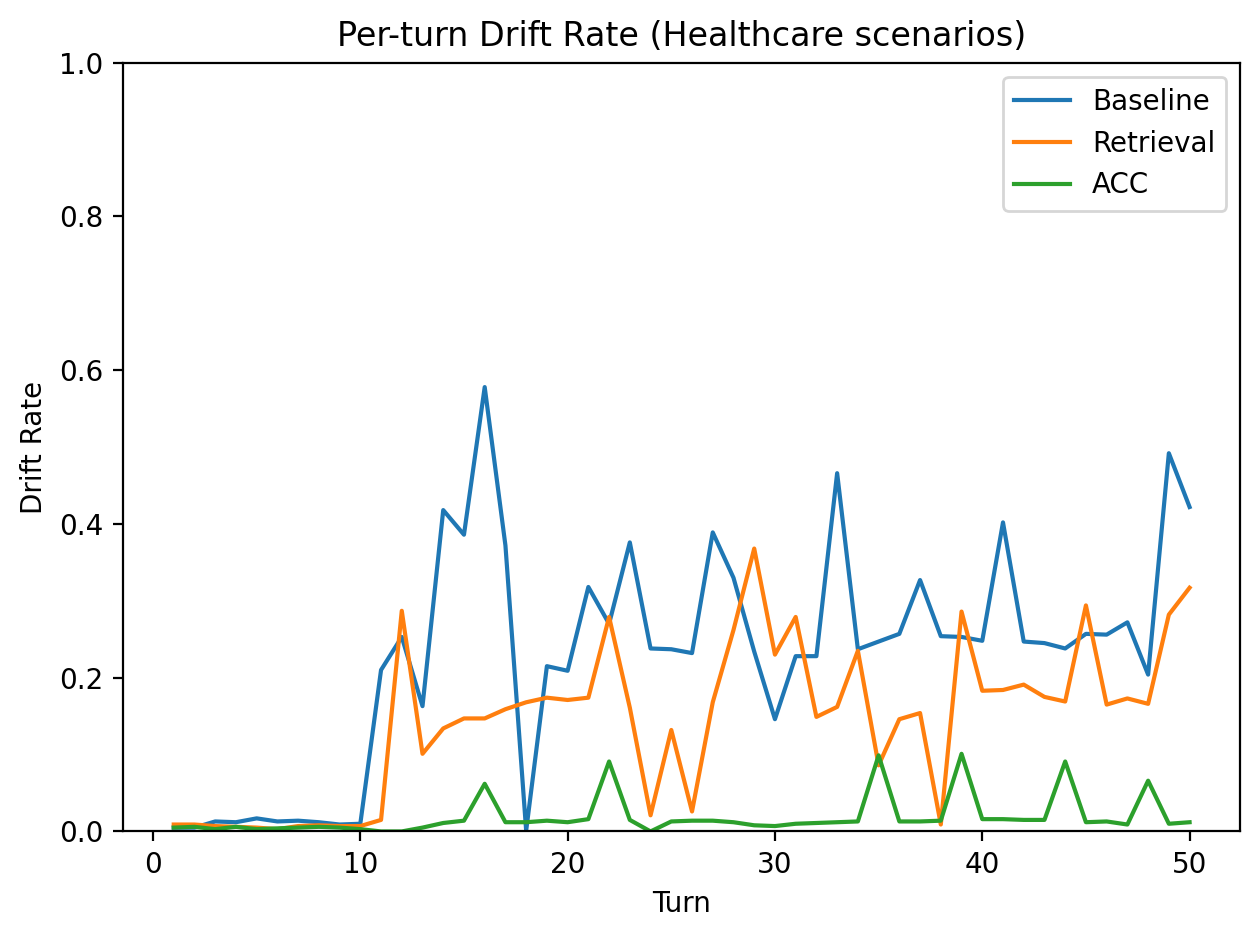}
    \caption{Healthcare: drift rate $D_t^{a}$.}
    \label{fig:health_drift}
\end{subfigure}

\vspace{0.7em}
% Row 3: Cybersecurity
\begin{subfigure}[t]{0.49\textwidth}
    \centering
    \includegraphics[width=\textwidth]{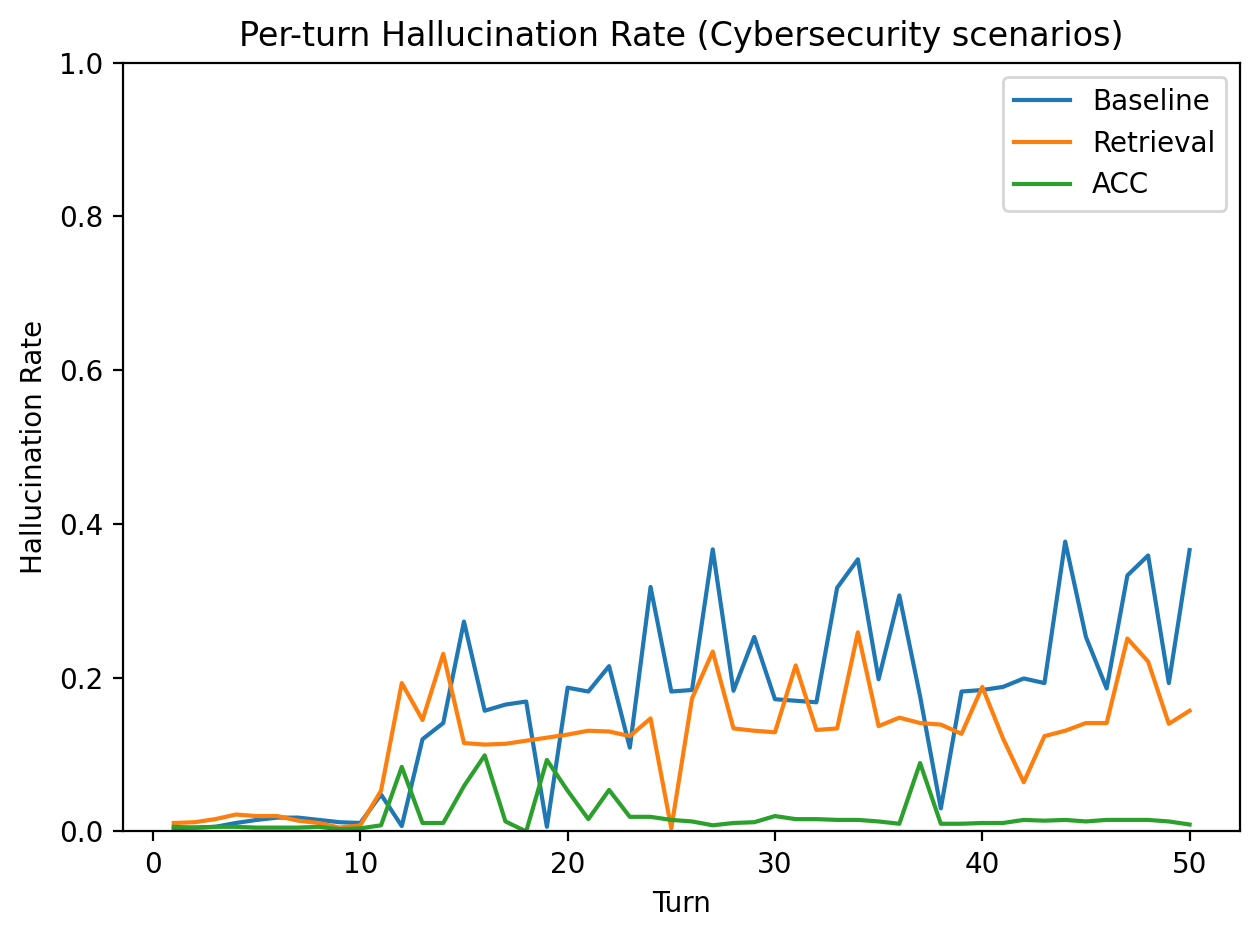}
    \caption{Cybersecurity: hallucination rate $H_t^{a}$.}
    \label{fig:cyber_hallucination}
\end{subfigure}
\hfill
\begin{subfigure}[t]{0.49\textwidth}
    \centering
    \includegraphics[width=\textwidth]{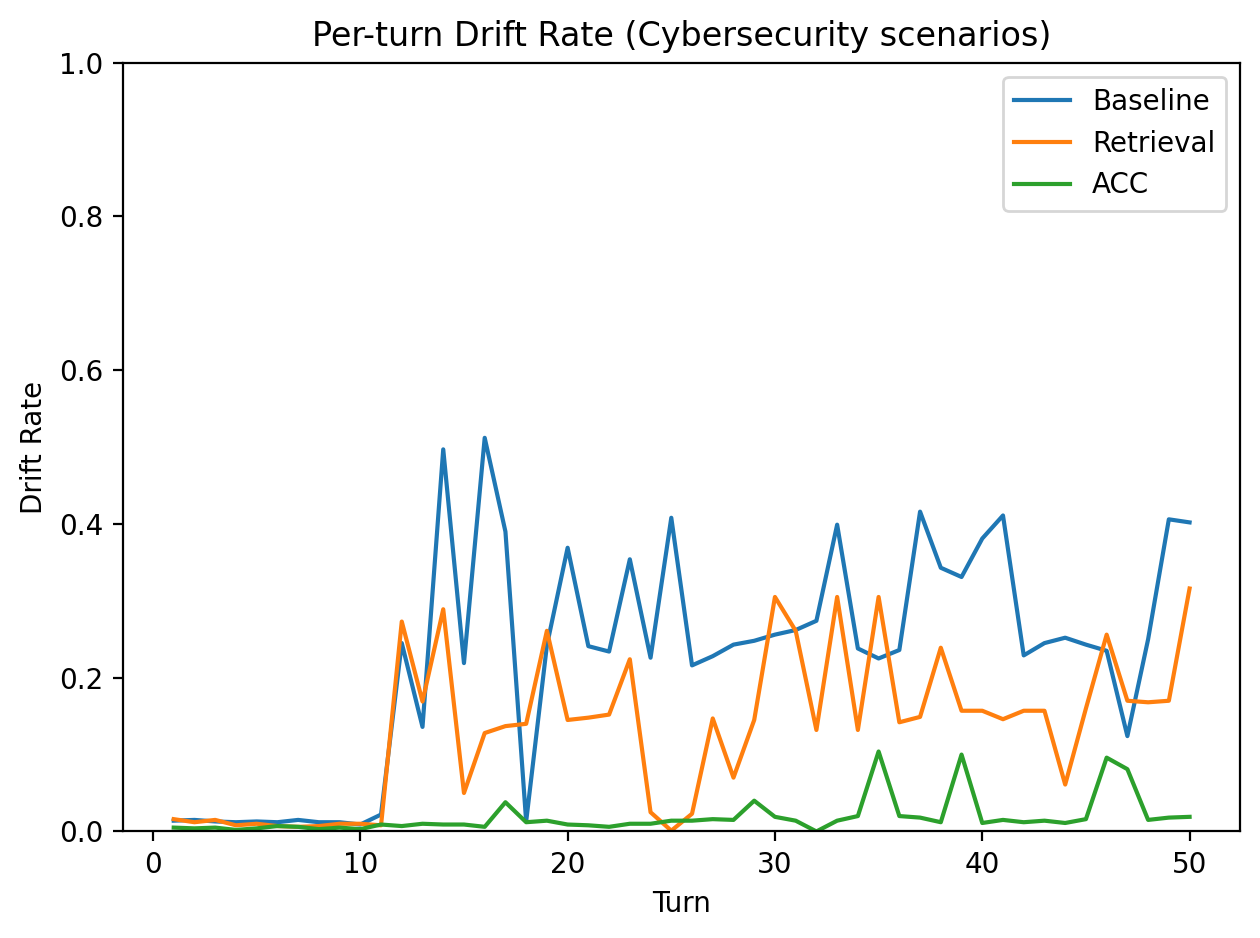}
    \caption{Cybersecurity: drift rate $D_t^{a}$.}
    \label{fig:cyber_drift}
\end{subfigure}

\caption{Turn-level memory reliability audits across domains over 50 turns. Left: hallucination rate $H_t^{a}$ from the judge claim audit grounded in canonical memory. Right: drift rate $D_t^{a}$ measuring constraint violations and missing requirements.}
\label{fig:audits_all_domains}
\end{figure}

Figure~\ref{fig:audits_all_domains} (right column) reports the turn-level drift rate $D_t^{a}$, which measures violations or omissions relative to active constraints in the judge canonical memory. Drift stays low early for all agents, then rises for baseline under conflicting requirements and strict formatting phases, and remains non-trivial for retrieval due to selection error (stale or competing artifacts). The ACC agent stays near zero with rare small spikes, demonstrating stronger constraint retention and more stable goal sustaining over the full 50-turn horizon

% ----------------------------
% Figures (ALL DOMAINS)
% ----------------------------

% --- Replace / eliminate the old plot: "average  metrics across domains .png"
% (Do NOT include it anymore. Use the figure below instead.)

\subsubsection{Why ACC outperforms}
Across the reported metrics, the results support a consistent explanation grounded in memory control. Transcript replay increases context length, so validated facts and constraints lose salience over time, enabling drift and hallucination carryover through repeated re-exposure. Retrieval bounds the amount of injected context but introduces selection error: when retrieval returns weakly related, stale, or injected artifacts, the agent conditions on the wrong information and errors propagate over turns. ACC addresses both mechanisms directly by (i) bounding the persistent conditioning signal by construction and (ii) separating artifact recall from state commitment, so recalled content only influences the next state through schema-governed compression. The combined effect is the observed cross-domain pattern: bounded memory growth together with higher outcome quality and substantially lower hallucination and drift rates.

\section{Conclusion}
\label{sec:conclusion}

This paper showed that multi-turn agent failures are often driven less by missing knowledge than by weak memory control. Transcript replay causes context to grow with turn count, reduces attention selectivity, and allows early errors to persist and reappear, which increases hallucination carryover and drift from established constraints. Retrieval-based memory bounds prompt length, but adds selection error: stale, conflicting, or injected artifacts can perturb the current task state and destabilize long-horizon behavior, which in our setting required restricting retrieval to three artifacts per turn to limit drift escalation.

We introduced the Agent Cognitive Compressor (ACC), a memory control mechanism that replaces accumulation with a bounded, schema-governed internal state. ACC separates artifact recall from state commitment and updates a single persistent variable, the Compressed Cognitive State (CCS), via controlled replacement rather than growth. This makes the write path explicit and auditable while keeping memory footprint bounded.

We also proposed an agent-judge-driven live evaluation framework that scores response quality and audits memory-driven anomalies at the turn level. Using a judge-maintained canonical memory, the framework computes grounded hallucination and drift rates alongside response-driven outcome metrics and direct measurements of memory footprint. Across domains including IT operations, cybersecurity response, healthcare operations, and finance workflows, ACC consistently maintained bounded memory and exhibited more stable multi-turn behavior with lower hallucination and drift than transcript replay and retrieval-based agents. The representative IT operations results highlight the core effect: ACC preserves constraint salience as horizons extend, while baseline and retrieval agents degrade under growing context or retrieval noise.

These findings support memory governance as a first-class requirement for reliable multi-turn agents. Future work will strengthen validation with targeted human audits, explore learned or task-adaptive CCS schemas, and study specialized compressor models and multi-agent extensions where state synchronization and shared constraints become central.

\bibliographystyle{plain} % Or another style like apa, ieee, etc.
\bibliography{science_template} % No need for the .bib extension

\end{document}